# Sequence co-evolution gives 3D contacts and structures of protein complexes

Thomas A. Hopf[1,2]*, Charlotta P.I. Schärfe[1,3]*, João P.G.L.M. Rodrigues[4]*, Anna G. Green[1], Chris Sander[5]#, Alexandre M.J.J. Bonvin[4]#, Debora S. Marks[1]#

[1] Department of Systems Biology, Harvard University, Boston, Massachusetts, USA; Lab: marks.hms.harvard.edu
[2] Bioinformatics and Computational Biology, Department of Informatics, Technische Universität München, Garching, Germany
[3] Applied Bioinformatics, Center for Bioinformatics, Quantitative Biology Center and Department of Computer Science, University of Tübingen, Germany
[4] Computational Structural Biology Group, Bijvoet Center for Biomolecular Research, Utrecht University, The Netherlands
[5] Computational Biology Center, Memorial Sloan Kettering Cancer Center, New York, NY, USA
* Joint first authors
# Correspondence to: EVcomplex@gmail.com

## Abstract

Protein-protein interactions are fundamental to many biological processes. Experimental screens have identified tens of thousands of interactions and structural biology has provided detailed functional insight for select 3D protein complexes. An alternative rich source of information about protein interactions is the evolutionary sequence record. Building on earlier work, we show that analysis of correlated evolutionary sequence changes across proteins identifies residues that are close in space with sufficient accuracy to determine the three-dimensional structure of the protein complexes. We evaluate prediction performance in blinded tests on 76 complexes of known 3D structure, predict protein-protein contacts in 32 complexes of unknown structure, and demonstrate how evolutionary couplings can be used to distinguish between interacting and non-interacting protein pairs in a large complex. With the current growth of sequence databases, we expect that the method can be generalized to genome-wide elucidation of protein-protein interaction networks and used for interaction predictions at residue resolution.

## Introduction

A large part of biological research is concerned with the identity, dynamics and specificity of protein interactions. There have been impressive advances in the three-dimensional (3D) structure determination of protein complexes which has been significantly extended by homology-inferred 3D models [1,2,3,4]. However, there is still little, or no, 3D information for ~80% of the currently known protein interactions in bacteria, yeast or human, amounting to at least ~30,000/~6000 incompletely characterized interactions in human and *E. coli,* respectively[2,5]. With the rapid rise in our knowledge of genetic variation at the sequence level, there is increased interest in linking sequence changes to changes in molecular interactions, but current experimental methods cannot match the increase in the demand for residue–level information of these interactions. One way to address the knowledge gap of protein interactions has been the use of hybrid, computational-experimental approaches that typically combine 3D structural information at varying resolutions, homology models and other methods [6], with force field-based approaches such as Rosetta Dock, residue cross-linking and data-driven approaches that incorporate various sources of biological information [1,7-16]. However, most of these approaches depend on the availability of prior knowledge and many biologically relevant systems remain out of reach, as additional experimental information is sparse (e.g. membrane proteins, transient interactions and large complexes). One promising computational approach is to use evolutionary analysis of amino acid co-variation to identify close residue contacts across protein interactions, which was first used 20 years ago[17,18], and subsequently used also to identify protein interactions [19,20]. Others have used some evolutionary information to improve a machine learning approach to developing docking potentials[21-23]. These previous approaches relied on a local model of co-evolution that is less likely to disentangle indirect and therefore incorrect correlations from the direct co-evolution, as has been described in work on residue-residue interactions in single proteins [24]. More recently, reports using a global model have been successful in identifying residue interactions from evolutionary covariation, for instance between histidine kinases and response regulators[25-27], and this approach has only recently been generalized and used to predict contacts between proteins in complexes of unknown structure, in an independent effort parallel to this work[28]. In principle, just a small number of key residue-residue contacts across a protein interface would allow computation of 3D models and provide a powerful, orthogonal approach to experiments.

Since the recent demonstration of the use of evolutionary couplings (ECs) between residues to determine the 3D structure of individual proteins [29-33], including integral membrane proteins [34,35], we reason that an evolutionary statistical approach such as EVcouplings[29] could be used to determine co-evolved residues *between* proteins. To assess this hypothesis we built an evaluation set based on all known binary protein interactions in *E. coli* that have 3D structures of the complex as recently summarized [5]. We develop a score for every predicted inter-protein residue pair based on the overall inter-protein EC score distributions resulting in accurate predictions for the majority of top ranked *inter*-protein EC pairs (inter-ECs) and sufficient to calculate accurate 3D models of the complexes in the docked subset, *Figure 1A*. This approach was then used to predict evolutionary couplings for 32 complexes of unknown 3D structures that have sufficient number of sequences, including previously published experimental support for our predicted unknown interactions between the a-, b- and c-subunit of ATP synthase.

## Results

We first investigated whether co-evolving residues between proteins are close in three dimensions by assessing blinded predictions of residue co-evolution against experimentally determined 3D complex structures. We follow this evaluation by then predicting co-evolved residue pairs of interacting proteins that have no known complex structure.

*Extension of the evolutionary couplings method to protein complexes*

To compute co-evolution across proteins, individual protein sequences must be aligned paired up with each other that are presumed to interact, or being tested to see if they interact. Without this condition, proteins could be paired together that do not in fact interact with each other and therefore detection of co-evolution would be compromised. Given that the evolutionary couplings method depends on large numbers of diverse sequences[34], some assumption must be made about which proteins interact with each other in homologous sequences in other species. Since it is challenging to know *a priori* whether particular interactions are conserved across many millions of years in thousands of different organisms, we use proximity of the two interacting partners on the genome as a proxy for this, with the goal of reducing incorrect pairings.

To assemble the broadest possible data sets to test the approach and make predictions we take all known interacting proteins assembled in a published dataset that contains ~3500 high-confidence protein interactions in *E. coli* [5]. After removing redundancy and requiring close genome distance between the pairs of proteins this results in 326 interactions, see Materials and methods (*Figure 1B, Figure 1 – figure supplement 1, Supplementary file 1 and 2*),

The paired sequences are concatenated and statistical co-evolution analysis is performed using EVcouplings [29,30,32], that applies a pseudolikelihood maximization (PLM) approximation to determine the interaction parameters in the underlying maximum entropy probability model [33,36], simultaneously generating both intra- and inter-EC scores for all pairs of residues within and across the protein pairs (*Figure 1A*). Evolutionary coupling calculations in previous work have indicated that this global probability model approach requires a minimum number of sequences in the alignment with at least 1 non-redundant sequence per residue [29-31,33,34]. Our current approach allows complexes with fewer available sequences to be assessed (minimum at 0.3 non-redundant sequences per residue) by using a new quality assessment score to assess the likelihood of the predicted contacts to be correct. The EVcomplex score is based on the knowledge that most pairs of residues are not coupled and true pair couplings are outliers in the high-scoring tail of the distribution (see Materials and methods, *Figure 2A and 2B, Figure 2 – figure supplement 1 and 2*).

The score can intuitively be understood as the distance from the noisy background of non-significant pair scores, normalized by the number of non-redundant sequences and the length of the protein (*Materials and methods, equations 1 and 2*). If the number of sequences per residue is not controlled for, there is a large bias in in the results, overestimating performance with low numbers of sequences (*Figure 2B and 2C*). The precise functional form of the correction for low numbers of sequences was chosen non-blindly after observing the dependencies in the test set.

*Blinded prediction of known complexes*

**Evolutionary covariation reveals inter-protein contacts**. Of the 329 interactions identified that are close on the *E. coli* genome, 76 have a sufficient number of alignable homologous sequences and known 3D structures either in *E. coli* or in other species. This set was used to test the inter-protein evolutionary coupling predictions (Supplementary file 1). The relationship between the EVcomplex score and the precision of the corresponding inter-protein ECs suggests that on average 74% (69%) of the predicted pairs with EVcomplex score greater than 0.8 will be accurate to within 10Å (8Å) of an experimental structure of the complex (*Figure 2C*). Most complexes have at least one inter-protein predicted contact above the selected score threshold of 0.8 (53/76 complexes). Three complexes have more than 20 predicted inter-protein residue contacts which are over 80% accurate, namely the histidine kinase and response regulator system (78 residue pairs), t-RNA synthetase (32 residue pairs and the vitamin B importer complex (21 residue pairs), with precision over 80% (complex numbers 330, 019, 130 respectively, *Figure 2D, Figure 2 – figure supplements 3-8, Supplementary file 1*).

We suggest that users of EVcomplex consider predicted contacts that lie below the threshold of 0.8 in the context of other biological knowledge, where available, or in comparison to other higher scoring contacts for the same complex. In this way additional true positive inter-residue contacts can be distinguished from false positives. For instance, the ethanolamine ammonia-lyase complex (complex 065) has only 3 predicted inter-protein residue pairs above the score threshold, but in fact has 5 additional correct pairs with EVcomplex scores slightly below the threshold of 0.8 which cluster with the 3 high-scoring contacts on the monomers, indicating that they are also correct.

Some of the high confidence inter-protein ECs in the test set are not close in 3D space when compared to their known 3D structures. These false positives may be a result of assumptions in the method that are not always correct. This includes (1) the assumption that the interaction between paired proteins is conserved across species and across paralogs, and (2) that truly co-evolved residues across proteins are indeed always close in 3D, which is not always the case. In addition, the complexes may also exist in alternative conformations that have not necessarily all been captured yet by crystal or NMR structures, for instance in the case of the large conformational changes of the BtuCDF complex [37].

**Docking is accurate with few pairs of predicted contacts.** To test whether the computed inter-protein ECs are sufficient for obtaining accurate 3D structures of the whole complex, we selected 15 diverse examples (with 5 or more inter-protein residue contacts) for docking (*Table 1, Figure 3, Figure 3 – figure supplement 1, Supplementary file 3*) with HADDOCK [14,38]. The docking procedure is fast and generates 100 3D models of each complex using all residue pairs with EVcomplex scores above the selection threshold. We additionally dock negative controls to assess the amount of information added to the docking protocol by evolutionary couplings (500 models per run, no constraints other than center of mass, see *Materials and methods*). The best models for all 15 complexes docked with evolutionary couplings have interface RMSDs under 6 Å, 12/15 have the best scoring model under 4Å and the top ranked models for 11/15 are under 5Å backbone interface RMSD compared to a crystal or NMR structure interface. Over 70% of the generated models are close to the experimental structures of the complexes (< 4Å backbone iRMSD), compared to less than 0.5% in the controls (and these were not high –ranked) (*Figure 3 – figure supplement 1, Supplementary file 3, Supplementary data*) Not surprisingly complexes that have the largest numbers of true positive predicted contacts perform the best when docking. For example, the ribosomal proteins RS3 and RS14 have 11 true positive inter-protein ECs and result in a top ranked model only 1.1 Å iRMSD from the reference structure. More surprisingly, other complexes with a lower proportion of true positive inter-protein contacts, such as Ubiquinol oxidase (6 out of 11) or the epsilon and gamma subunits of ATP synthase (8 out of 15) also produced accurate predicted complexes, with an iRMSD of 1.8 and 1.4 Å respectively. The docking experiments therefore demonstrate that inter-protein ECs, even in the presence of incorrect predictions, can be sufficient to give accurate 3D models of protein complexes, but more work will be needed to quantify the likelihood of successful docking from the predicted contacts.

**Conserved residue networks provide evidence of functional constraints.** The top 10 inter-EC pairs between MetI and MetN are accurate to within 8Å in the MetNI complex (PDB: 3tui [39]), resulting in an average 1.4 Å iRMSD from the crystal structure for all 100 computed 3D models (*Table 1, Supplementary file 3 and Supplementary data).* The top 3 inter-EC residue pairs (K136-E108, A128-L105, and E74-R124, MetI-MetN respectively) constitute a residue network coupling the ATP binding pocket of MetN to the membrane transporter MetI. This network calculated from the sequence alignment corresponds to residues identified experimentally that couple ATP hydrolysis to the open and closed conformations of the MetI dimer [39] (*Figure 4A*). The vitamin B12 transporter (BtuC) belongs to a different structural class of ABC transporters, but also uses ATP hydrolysis via an interacting ATPase (BtuD). The top 5 inter-ECs co-locate the L-loop of BtuC close to the Q-loop ATP-binding domain of the ATPase, hence coupling the transporter with the ATP hydrolysis state in an analogous way to MetI-MetN. The identification of these coupled residues across the different subunits suggests that EVcomplex identifies not only residues close in space, but also particular pairs that are constrained by the transporter function of these complexes [39,40].

The ATP synthase ε and γ subunit complex provides a challenge to our approach, since the ε subunit can take different positions relative to the γ subunit, executing the auto-inhibition of the enzyme by dramatic conformational changes [41]. In a real-world scenario, where we might not know this *a priori*, there may be conflicting constraints in the evolutionary record corresponding to the different positions of the flexible portion of ε subunit. EVcomplex accurately predicts 6 of the top 10 inter-EC pairs (within 8Å in the crystal structure 1fs0[42] or 3oaa[41]), with the top 2 inter-ECs εA45-γL215 and εA40-γL207 providing contact between the subunits along an inter-protein beta sheet. The location of the C-terminal helices of the ε subunit is significantly different across 3 crystal structures (PDB IDs: 1fs0[42], 1aqt [43], 3oaa [41]). The top ranked intra-ECs support the conformation seen in 1aqt, with the C-terminal helices packed in an antiparallel manner and tucked against the N-terminal beta barrel (*Figure 4B,* green circles) and do not contain a high ranked evolutionary trace for the extended helical contact to the γ subunit seen in 1fs0 or 3oaa (*Figure 4B*, grey box). Docking with the top inter-ECs results in models with 1.4 Å backbone iRMSDs to the crystal structure for the interface between the N-terminal domain of the ε subunit and the γ subunit (*Table 1, Supplementary file 4*). εD82 and γR222 connect the ε-subunit via a network of 3 high-scoring intra-ECs between the N- and C-terminal helices to the core of the F1 ATP synthase. In summary, these examples suggest that inter-protein evolutionary couplings can provide residue relationships across the proteins that could aid identification of functional coupling pathways, in addition to obtaining 3D models of the complex.

*De novo prediction of unknown complexes.*

**Prediction of interactions for 32 protein pairs with high-scoring evolutionary couplings**. A total of 82 protein complexes with unknown 3D structure of the interaction that satisfy the conditions for the current approach, i.e. have sufficient sequences and are close in all genomes, were predicted using EVcomplex (all residue – residue inter protein evolutionary couplings scores are available in *Supplementary data*). 32 of these have high EVcomplex scores with at least one predicted contact (*Figure 5, Figure 5 – figure supplement 1 and 2, and Supplementary file 4)*. Analysis of the inter-EC predictions for known 3D complex structures shows that protein pairs with more high-scoring ECs (EVcomplex score > 0.8) have a higher proportion of true positives (*Figure 2D*). Hence, the protein complexes in the set of unknown structures with more high-scoring inter-ECs are the most likely to have predicted ECs that indicate residue pairs close in 3D (column Q, *Supplementary file 2,* the exact pairs can be found in *Supplementary file 4)*. Three examples of predictions with multiple high-scoring inter-ECs include MetQ-MetI, UmuD-UmuC and DinJ-YafQ. The top 15 inter-ECs between MetQ and MetI are from one interface of MetQ to the MetI periplasmic loops, or the periplasmic end of the helices, consistent with the known binding of MetQ to MetI in the periplasm.

The UmuD and UmuC complex is induced in the stress/SOS response facilitating the cleavage of UmuD to UmuD' (between C24 and G25) to form UmuD'$_2$ which then interacts with UmuC (DNA polymerase V) in order to copy damaged DNA[44]. The truncated dimer form (UmuD'$_2$) has at least two contrasting conformations where the N-terminal arm is placed on opposite sides of the dimer in one conformation or in close proximity in the alternative (*Figure 5 – figure supplement 3)*. For 6/7 ECs above the score threshold, residues in UmuD predicted to interface with UmuC are co-located on one face of the dimer. Two residues (Y33, I38) are located in the N-terminal arm of UmuD that, after cleavage of the 24 N-terminal amino acids, may become available for binding UmuC. Since UmuD switches functions after this cleavage and can then bind UmuC, these inter ECs may identify the critical residues for translesion synthesis function[44]. Although the ECs from this UmuD arm to UmuC involve residues in two separate domains of UmuC (S 415 and Y 74), intra-monomer evolutionary couplings predict that these residues are close in UmuC (*Figure 5 - figure supplement 3A, black rectangles*). The relative positions of the contacting residues within each monomer therefore support the plausibility of the accuracy of the interaction interface.

Whilst this manuscript was in review, the 3D structure of the previously unsolved biofilm toxin/antitoxin DinJ-YafQ complex was published (PDB: 3mlo[45]), showing the intertwining of subunits in a heterotetrameric complex. 17/19 predicted EC residue pairs are within 8 Å in this 3D structure (*Supplementary file 4 and Supplementary data*). In general, the agreement between our *de novo* predicted inter-protein ECs with available experimental data serves as a measure of confidence for the predicted residue pair interactions, and suggests that EVcomplex can be used to reveal 3D structural details of yet unsolved protein complexes given sufficient evolutionary information.

**EVcomplex predicts interacting protein pairs in a large complex.** To investigate whether the EVcomplex score can also distinguish between interacting and non-interacting pairs of proteins, we use the *E. coli* ATP synthase complex as a test case. The ATP synthase structure is of wide biological interest (reviewed in [46]) with a remarkable 3D structural arrangement, but completion of all aspects of the 3D structure has remained experimentally challenging [47] (*Figure 6A*). As a demonstration exercise, we calculated evolutionary couplings for all 28 possible pair combinations of different ATP synthase subunits (centered around the *E. coli* ATP synthase) and transformed the ECs into EVcomplex scores for all inter-protein residue pairs (experimentally determined stoichiometry: $\alpha_3\beta_3\gamma\delta\epsilon ab_2c_{10}$, *Supplementary file 5 and Supplementary data*). Using the default EVcomplex score threshold of 0.8 to discriminate between interacting and non-interacting pairs of subunits, 24 of the 28 possible interactions between the subunits are correctly classified as interacting or non-interacting. The four incorrect predictions (namely: ε and c, γ and c, ε and β, b and β, for which there is some experimental evidence) are not identified as interacting using the 0.8 EVcomplex threshold. Choosing a threshold lower than 0.8 does identify 2 of these as interacting but also introduces new false positives. The ε and β interaction in the crystal structure 3oaa[41] is a special case in that it involves a highly extended conformation of the last two helices of the ε subunit that reach up into the enzyme making contacts with the β subunit. The false negative EVcomplex score for this pair could be a result of the transience of their interaction or reflect a more general problem of lack of conservation of this interaction across the aligned proteins from different species. In total 80% of the interacting residue pairs in the known 3D structure parts of the synthase complex (7 pairs of subunits) are correctly predicted (threshold: 10Å minimum atom distance between two residues). This exercise of prediction of presence or absence of interaction between any two proteins indicates the potential of the EVcomplex method in helping elucidate protein-protein interaction networks from evolutionary sequence co-variation and identify interacting subunits of large macromolecular complexes.

**EVcomplex predicts details of subunit interactions in ATP synthase**. While much of the 3D structure of ATP synthase is known[46], the details of interactions between the a- b-, and c-subunits have not yet been determined by crystallography. We analyse the details in these interactions, as the EVcomplex scores between these subunits are substantial (*Figure 6B*). We are fortunately able to provide a missing piece for this analysis, the unknown structure of the membrane-integral penta-helical a-subunit, using our previously described method for *de novo* 3D structure prediction of alpha-helical transmembrane proteins [34]. To our knowledge there are no experimentally determined atomic resolution structures of the a-subunit of ATP synthase. A 3D model of the a-subunit is from 1999 (1c17[48]) and was computed using five helical–helical interactions that were inferred from second suppressor mutation experiments, and then imposed as distance restraints for TMH2-5, revealing a four helical bundle (with no information for TMH1). Later, cross-linking experiments [49] identified contacting residues from all pairs of helical combinations of TM2-TM5 (6 pairs), supporting the earlier 4 helical bundle topology. 7 of the 8 cross-linked pairs are either exactly the same pair (L120-I246) or adjacent to many pairs in the top L intra a-subunit evolutionary couplings (ECs).

In fact, the helix packing arrangement in the predicted structure of the a-subunit is consistent with the topology suggested on the basis of crosslinking studies [50-52], including the lack of contacts for transmembrane helix 1 with the other 4 helices (*Supplementary data*).

The top inter-protein EC pair between subunits a and b, aK74–bE34, coincides with experimental crosslinking evidence of the interaction of aK74 with the b-subunit and the position of E34 of the b subunit emerging from the membrane on the cytoplasmic side[50,51]. Indeed, 6 of the 13 high score ECs are in the same region as the experimental crosslinks, for instance between the cytoplasmic loop between the first two helices of the a-subunit and the b-subunit helix as it emerges from the membrane bilayer [53], a239V in TM helix 5 and bL16 (*Figure 6C, Figure 6 – figure supplement 1, Supplementary file 6*). Additionally, the top EC between the a- and c-subunits (aG213 – cM65) lies close to the functionally critical aR210–cD61 interaction [54] on the same helical faces of the respective subunits (*Figure 6C)*. This prediction of missing aspects of subunit interactions may help in the design of targeted experiments to complete the understanding of the intricate molecular mechanism of the ATP synthase complex.

## Discussion

A primary limitation of our current approach is its dependence on the availability of a large number of evolutionarily related sequences. If a protein interaction is conserved across enough sequenced genomes, using a single pair per genome can give accurate predictions of the interacting residues. However, if the protein pair is present in limited taxonomic branches, there may be insufficient sequences at any given time to make confident predictions. A solution to this could be to include multiple paralogs of the interacting proteins from each genome, but this requires correct pairing of the interaction partners, which is in general hard to ascertain. In addition, details of interactions may have diverged for paralogous pairs. Hence, in this current version of the method we have imposed a genome distance requirement across all genomes for all homolog pairs in order to be less sensitive to these complications.

As the need to use genome proximity to pair sequences becomes less important with the increasing availability of genome sequences, there will be a dramatic increase in the number of interactions that can be inferred from evolutionary couplings, including those unique to eukaryotes. With currently available sequences (May 2014 release of the UniProt database), EVcomplex is able to provide information for about $1/10^{th}$ of the known 3000 protein interactions in the *E. coli* genome. Once there are ~10,000 bacterial genome sequences of sufficient diversity, one would have enough information to test each potentially interacting pair of homologs for evidence of interaction and, given sufficiently strong evolutionary couplings, infer the 3D structure of each protein-protein pair, as well as of complexes with more than two proteins. For any set of species, e.g., vertebrates or mammals, one can imagine guiding sequencing efforts to optimize species diversity to facilitate the extraction of evolutionary couplings. This can open the doors for more comprehensive and more rapid determination of approximate 3D structures of proteins and protein complexes, as well as for the elucidation in molecular detail of the most strongly evolutionarily constrained interactions, pointing to functional interactions.

Determining the three-dimensional models of complexes from the predicted contacts was successful in many of the tested instances. Using minimal computing resources and a small number of inter-EC-derived contacts, low interface positional RMSDs relative to experimental structures can be achieved. However, a significant number of proteins exist as homomultimers within larger complexes. To determine models of these complexes one must deconvolute homomultimeric inter-ECs from the intra-protein signal, which is an important technical challenge for future work.

The analysis of subunit interactions in ATP synthase in this work is a "proof of principle" study showing that methods such as EVcomplex can determine which proteins interact with each other at the same time as specific residue pair couplings across the proteins (as also shown in the work by the Baker lab on ribosomal protein interactions[28]). Understanding the networks of protein interactions is of critical interest in eukaryotic systems, such as networks of protein kinases, GPCRs, or PDZ domain proteins. An understanding of the distributions of interaction specificities is of high interest to many fields. Although we do not know how well our evolutionary coupling approach will handle less obligate interactions, results on the two-component signalling system (histidine kinase/response regulator) both here and in other work[25,26] suggest optimism.

The approximately scale-free EVcomplex score is a heuristic based on the distribution of raw EC scores from the statistical model, their dependence on sequence alignment depth and the length of the concatenated sequences. The score provides a simple way of accounting for these dependencies such that a uniform threshold, say 0.8, can be used for any protein pair with the expectation of reasonably accurate predictions. Since cutoff thresholds can be useful but overly sharp, we recommend investigating predicted contacts below the threshold used in this work, especially where there is independent biological knowledge to validate the predictions.

The work presented here is in anticipation of a genome-wide exploration and, as a proof of principle, shows the accurate prediction of inter-protein contacts in many cases and their utility for the computation of 3D structures across diverse complex interfaces. As with single protein (intra-EC) predictions, evolutionarily conserved conformational flexibility and oligomerization can result in more than one set of contacts that must be de-convoluted. Can evolutionary information help to predict the details and extent for each complex? A key challenge will be the development of algorithms that can disentangle evolutionary signals caused by alternative conformations of single complexes, alternative conformations of homologous complexes, and effectively deal with false positive signals. Taken together, these issues highlight fruitful areas for future development of evolutionary coupling methods.

Despite conditions for the successful *de novo* calculation of co-evolved residues, the method described here may accelerate the exploration of the protein-protein interaction world and the determination of protein complexes on a genome-wide scale at residue level resolution. The use of co-evolutionary analysis in computational models to determine protein specificity and promiscuity, co-evolutionary dynamics and functional drift will open up exciting future research questions.

## Materials and methods

*Selection of interacting protein pairs for co-evolution calculation.*

The candidate set of complexes for testing and *de novo* prediction was derived starting from a dataset of binary protein-protein interactions in *E. coli* including yeast two-hybrid experiments, literature-curated interactions and 3D complex structures in the PDB [5]. Three complexes not contained in the list were added based on our analysis of other subunits in the same complex, namely BtuC/BtuF, MetI/MetQ, and the interaction between ATP synthase subunits a and b. Since our algorithm for concatenating multiple sequence pairs per species assumes the proximity of the interacting proteins on the respective genomes of each species (see below), we excluded any complex with a gene distance > 20 from further analysis. The gene distance is calculated as the number of genes between the interacting partners based on an ordered list of genes in the *E. coli* genome obtained from the UniProt database. The resulting list of pairs (~ 350) was then filtered for pseudo-homomultimeric complexes based on the identification of Pfam domains in the interacting proteins (330). All remaining complexes with a known 3D structure (as summarized in [5]) or a homologous interacting 3D structure (93) (identified by intersecting the results of HMMER searches against the PDB for both monomers) were used for evaluating the method, while complexes without known structure (236) were assigned to the *de novo* prediction set (*Figure 1 – figure supplement 1*). The set with protein complexes of known 3D structure was further filtered for structures that only cover fragments (< 30 amino acids) of one or both of the monomers and structures with very low resolution (> 5Å), which led to the re-assignment of Ribonucleoside-diphosphate reductase 1 (complex_002), Type I restriction-modification enzyme EcoKI (complex_012), RpoC/RpoB (complex_041), RL11/Rl7 (complex_165), the ribosome with SecY (complex_226, complex_250, and complex_255), and RS3/RS (complex_254) to the set of unknown complexes. Large proteins were run with the specific interacting domains informed by the known 3D structure, when the full sequence was too large for the number of retrieved sequences, (for domain annotation see Supplementary data.)
This set could serve as a benchmark set for future development efforts in the community.

*Multiple sequence alignments.*

Each protein from all pairs in our dataset was used to generate a multiple sequence alignment (MSA) using jackhmmer [55] to search the UniProt database[56] with 5 iterations. To obtain alignments of consistent evolutionary depths across all the proteins, a bit score threshold of 0.5 * monomer sequence length was chosen as homolog inclusion criterion (-incdomT parameter), rather than a fixed *E*-value threshold which selects for different degrees of evolutionary divergence based on the length of the input sequence.

In order to calculate co-evolved residues across different proteins, the interacting pairs of sequences in each species need to be matched. Here, we assume that proteins in close proximity on the genome, e.g., on the same operon, are more likely to interact, as in the methods used previously matching histidine kinase and response regulator interacting pairs [25,26] (*Supplementary* *data*). We retrieved the genomic locations of proteins in the alignments and concatenated pairs following 2 rules: (i) The CDS of each concatenated protein pair must be located on the same genomic contig (using ENA [57] for mapping), and (ii) each pair must be the closest to one another on the genome, when compared to all other possible pairings in the same species. The concatenated sequence pairs were filtered based on the distribution of genomic distances to exclude outlier pairs with high genomic distances of more than 10k nucleotides (*Supplementary data*). Alignment members were clustered together and reweighted if 80% or more of their residues were identical (thus implicitly removing duplicate sequences from the alignment). *Supplementary file 1 and 2* report the total number of concatenated sequences, the lengths, and the effective number of sequences remaining after down-weighting in the evaluation and de novo prediction set, respectively.

*Computation of evolutionary couplings.*

Inter- and intra-ECs were calculated on the alignment of concatenated sequences using a global probability model of sequence co-evolution, adapted from the method for single proteins[29,30,34] using a pseudo-likelihood maximization (PLM) [36] rather than mean field approximation to calculate the coupling parameters. Columns in the alignment that contain more than 80% gaps were excluded and the weight of each sequence was adjusted to represent its cluster size in the alignment thus reducing the influence of identical or near-identical sequences in the calculation. For the evaluation set we can then compare the predicted ECs for both within and between the protein/domains to the crystal structures of the complexes (for contact maps and all EC scores, see *Supplementary data*).

*Definition of a scale-free score for the assessment of interactions.*

In order to estimate the accuracy of the EC prediction we evaluate the calculated inter-ECs based on the following observations: (1) most pairs of positions in an alignment are not coupled, i.e. have an EC score close to zero, and tend to be distant in the 3D structure; (2) the background distribution of EC scores between non-coupled positions is approximately symmetric around a zero mean; and (3) higher-scoring positive score outliers capture 3D proximity more accurately than lower-scoring outliers (see also *Figure 2*). The width of the (symmetric) background EC score distribution can be approximated using the absolute value of the minimal inter-EC score. The more a positive EC score exceeds the noise level of background coupling, the more likely it is to reflect true co-evolution between the coupled sites. For each inter-protein pair of sites i and j with pair coupling strength $EC_{inter}(i, j)$, we therefore calculate a raw reliability score ('pair coupling score ratio', *Figure 2B*) defined by

$$Q_{\text{inter}}^{\text{raw}}(i,j) = \frac{EC_{\text{inter}}(i,j)}{\left|\min_{i,j}\left(EC_{\text{inter}}(i,j)\right)\right|} \qquad (1)$$

Since the accuracy of evolutionary couplings critically depends both on the number and diversity of sequences in the input alignment and the size of the statistical inference problem [29-31] we incorporate a normalization factor to make the raw reliability score comparable across different protein pairs. The normalized EVcomplex score is defined as

$$\text{EVcomplex-Score}(i,j) = \frac{Q_{\text{inter}}^{\text{raw}}(i,j)}{1 + \left(\frac{N_{\text{eff}}}{L}\right)^{-\frac{1}{2}}} \qquad (2)$$

where $N_{\text{eff}}$ is the effective number of sequences in the alignment after redundancy reduction, and L (total number of residues) is the length of the concatenated alignment. Previous work on single proteins has shown that the method requires a sufficient number of sequences in the alignment to be statistically meaningful. We thus filter for sequence sufficiency requiring $N_{\text{eff}}/L > 0.3$ (*Table 1, Supplementary files 1 and 2*). Predictions of coupled residues in the evaluation set were evaluated against their residue distances in known structures of protein pairs [5] (see *Supplementary file 7*) in order to determine the precision of the method.

To interpret the EVcomplex prediction of interaction between subunits a and b of the ATP synthase as well as UmuC and UmuD, individual monomer models were built *de novo* for the structurally unsolved subunit-a of ATP synthase and UmuC using the EVfold pipeline as previously published[29,34]. In both cases coupling parameters were calculated using PLM [36] and sequences were clustered and weighted at 90% sequence identity (the resulting models are provided in *Supplementary data*).

*Prediction of interactions in a set of subunits.*

Following this same protocol EVcomplex scores were calculated for all possible 28 combinations of the 8 *E. coli* ATP synthase $F_0$ and $F_1$ subunits. Since we want to compare the computational predictions to some 'ground truth', as with the complexes for the rest of the manuscript, we used known 3D structures of the ATP synthase complex to assign whether or not the subunits interact (3oaa, 1fs0, 2a7u *Supplementary file 7*). Since we are also determining whether the subunits interact, not necessarily knowing full atomic detail residue interactions, we included subunit interactions that have been inferred from cryo-EM, crosslinking or other experiments, but do not necessarily have a crystal structure. These are represented as solid blue boxes, if the interaction is well established[53,58-60], or crosshatched blue if there is a lack of consensus in the community, left panel *Figure 6B*.

For each possible interaction the EVcomplex score of the highest ranked inter-EC was considered as a proxy for the likelihood of interaction. Pairs with scores above 0.8 are considered likely to interact, between 0.75 and 0.8 weakly predicted, while interactions with scores below 0.75 are rejected as possible complexes, blue boxes, blue crosshatched and white respectively, right panel *Figure 6B and Supplementary data.*

***Computation of 3D structure of complexes.***

A diverse set of 15 complexes was chosen from the 22 in the evaluation set that had at least 5 couplings above a complex score of 0.8 and were subsequently docked (*Supplementary file 3*). Proteins that have been crystallized together in a complex could bias the results of the docking, as they have complementary positions of the surface side chains. Therefore, where possible we used complexes that had a solved 3D structure of the unbound monomer, namely GcsH/GcsT, CyoA, FimC, DhaL, AtpE, PtqA/PtqB, RS10 and HK/RR, and in all other cases the side chains of the monomers were randomized either by using SCWRL4 [61] or restrained minimization with Schrodinger Protein Preparation Wizard[62] before docking. For ubiquinol oxidase (complex_054) the unbound structure of subunit 2 (CyoA) only covers the COX2 domain. In this case docking was performed using this unbound structure plus an additional run using the bound complex structure with perturbed side chains.

We used HADDOCK [14], a widely used docking program based on ARIA [63] and the CNS software [64] (Crystallography and NMR System), to dock the monomers for each protein pair with all inter-ECs with an EVcomplex score of 0.8 or above implemented as distance restraints on the α–carbon atoms of the backbone.

Each docking calculation starts with a rigid-body energy minimization, followed by semi-flexible refinement in torsion angle space, and ends with further refinement of the models in explicit solvent (water). 500/100/100 models generated for each of the 3 steps, respectively. All other parameters were left as the default values in the HADDOCK protocol. Each protein complex was run using predicted ECs as unambiguous distance restraints on the Cα atoms ($d_{eff}$ 5Å, upper bound 2Å, lower bound 2Å; input files available in Supplementary data). As a negative control, each protein complex was also docked using center of mass restraints (*ab initio* docking mode of HADDOCK) [38] alone and in the case of the controls generating 10000/500/500 models.

Each of the generated models is scored using a weighted sum of electrostatic ($E_{elec}$) and van der Waals ($E_{vdw}$) energies complemented by an empirical desolvation energy term ($E_{desolv}$)[65]. The distance restraint energy term was explicitly removed from the equation in the last iteration (Edist3 = 0.0) to enable comparison of the scores between the runs that used a different number of ECs as distance restraints.

***Comparison of predicted to experimental structures.***

All computed models in the docked set were compared to the cognate crystal structures by the RMSD of all backbone atoms at the interface of the complex using ProFit v.3.1 (http://www.bioinf.org.uk/software/profit/). The interface is defined as the set of all residues that contain any atom < 6 Å away from any atom of the complex partner. For the AtpE-AtpG complex we excluded the 2 C-terminal helices of AtpE as these helices are mobile and take many different positions relative to other ATP synthase subunits [41]. Similarly, since the DHp domain of histidine kinases can take different positions relative to the CA domain, the HK-RR complex was compared over the interface between the DHp domain alone and the response regulator partner. In the case of the unbound ubiquinol oxidase docking results, only the interface between COX2 in subunit 2 and subunit 1 was considered. Accuracy of the computed models with EC restraints were compared with computed models with center of mass restraints alone (negative controls), *Figure 3 – figure supplement 1, Supplementary file 3*).

Data analysis was conducted primarily using IPython notebooks[66]. A webserver and all data is made EVcomplex.org.

Table Legends

Table 1. EVcomplex predictions and docking results for 15 protein complexes

| Complex Name | Subunits | EVcomplex contacts | | | Docking quality (iRMSD) | |
| --- | --- | --- | --- | --- | --- | --- |
| | | Seqs[a] | ECs[b] | TP rate[c] | Top ranked model[d] | Best model[e] |
| Carbamoyl-phosphate synthase | CarB:CarA | 2.3 | 17 | 0.88 | 1.9 | 1.9 |
| Aminomethyltransferase/ Glycine cleavage system H protein | GcsH:GcsT | 2.9 | 5 | 0.2 | 5.4 | 5.4 |
| Histidine kinase/ response regulator | KdpD:CheY (*T. maritima*) | 95.4 | 78 | 0.72 | 2.1 | 2.0 |
| Ubiquinol oxidase | CyoB:CyoA | 1.0 | 11 | 0.55 | 1.8 | 1.2 |
| Outer membrane usher protein/ Chaperone protein | FimD:FimC | 3.6 | 6 | 0.83 | 3.2 | 3.0 |
| Molybdopterin synthase | MoaD:MoaE | 3.6 | 8 | 1.0 | 4.4 | 4.1 |
| Methionine transporter complex | MetN:MetI | 1.9 | 14 | 0.86 | 1.5 | 1.2 |
| Dihydroxyacetone kinase | DhaL:DhaK | 1.4 | 12 | 0.42 | 6.7 | 2.4 |
| Vitamin B12 uptake system | BtuC:BtuF | 3.2 | 5 | 0.6 | 2.8 | 2.8 |
| Vitamin B12 uptake system | BtuC:BtuD | 9.8 | 21 | 0.88 | 1.1 | 0.9 |
| ATP synthase γ and ε subunits | AtpE:AtpG | 2.9 | 15 | 0.53 | 1.4 | 1.4 |
| IIA-IIB complex of the N,N'-diacetylchitobiose (Chb) transporter | PtqA:PtqB | 3.1 | 5 | 0.2 | 7.2 | 5.5 |
| 30 S Ribosomal proteins | RS3:RS14 | 1.4 | 11 | 0.91 | 1.1 | 1.1 |
| Succinatequinone oxido-reductase flavoprotein/ iron-sulfur subunits | SdhB:SdhA | 3.0 | 8 | 0.62 | 1.4 | 1.4 |
| 30 S Ribosomal proteins | RS10:RS14 | 1.2 | 6 | 1.0 | 5.3 | 2.5 |

[a]Number of non-redundant sequences in concatenated alignment normalized by alignment length, [b]inter-ECs with EVcomplex score ≥ 0.8, [c]True Positive rate for inter ECs above score threshold, [d]iRMSD positional deviation of model from known structure, for docked model with best HADDOCK score, [e]lowest iRMSD observed across all models

Figure Legends

**Figure 1. Figure 1. Co-evolution of residues across protein complexes from the evolutionary sequence record**. (A) Evolutionary pressure to maintain protein-protein interactions leads to the coevolution of residues between interacting proteins in a complex. By analyzing patterns of amino acid co-variation in an alignment of putatively interacting homologous proteins (left), evolutionary couplings between coevolving inter-protein residue pairs can be identified (middle). By defining distance restraints on these pairs, the 3D structure of the protein complex can be inferred using docking software (right). (B) Distribution of E. coli protein complexes of known and unknown 3D structure where both subunits are close on the bacterial genome (left), allowing sequence pair matching by genomic distance. For a subset of these complexes, sufficient sequence information is available for evolutionary couplings analysis (dark blue bars). As more genomic information is created through on-going sequencing efforts, larger fractions of the E. coli interactome become accessible for EVComplex (right). A detailed version of the workflow used to calculate all E. coli complexes currently for which there is currently enough sequence information is shown in Figure1 - figure supplement 1.

**Figure 2. Evolutionary couplings capture interacting residues in protein complexes**. (A) Inter- and Intra-EC pairs with high coupling scores largely correspond to proximal pairs in 3D, but only if they lie above the background level of the coupling score distribution. To estimate this background noise a symmetric range around 0 is considered with the width being defined by the minimum inter-EC score. For the protein complexes in the evaluation set this distribution is compared to the distance in the known 3D structure of the complex that is shown here for the methionine transporter complex, MetNI. (Plots for all complexes in the evaluation set are shown in Figure 2 - figure supplement 1 and 2). (B) A larger distance from the background noise (ratio of EC score over background noise line) gives more accurate contacts. Additionally, the higher the number of sequences in the alignment the more reliable the inferred coupling pairs are which then reduces the required distance from noise (different shades of blue). Residue pairs with an 8Å minimum atom distance between the residues are defined as true positive contacts, and precision = TP/(TP+FP). The plot is limited to range (0,3) which excludes the histidine kinase – response regulator complex (HK-RR) – a single outlier with extremely high number of sequences. (C) To allow the comparison across protein complexes and to estimate the average inter-EC precision for a given score threshold independent of sequence numbers, the raw couplings score is normalized for the number of sequences in the alignment, the EVcomplex score. In this work, inter-ECs with a score ≥ 0.8 are used. Note: the shown figure is cut off at score of 2 in order to zoom in on the phase change region and the high sequence coverage outlier HK-RR is excluded. (D) For complexes in the benchmark set, inter-EC pairs with EVcomplex score ≥ 0.8 give predictions of interacting residue pairs between the complex subunits to varying accuracy (8Å TP distance cutoff). All predicted interacting residues for complexes in the benchmark set that had at least one inter-EC above 0.8 are shown as contact maps in Figure 2 – figure supplement 3-8.

**Figure 3. Blinded prediction of evolutionary couplings between complex subunits with known 3D structure**. Inter-ECs with EVcomplex score ≥ 0.8 on a selection of benchmark complexes (monomer subunits in green and blue, inter ECs in red, pairs closer than 8Å by solid red lines, dashed otherwise). The predicted inter-ECs for these ten complexes were then used to create full 3D models of the complex using protein-protein docking For the fifteen complexes for which also 3D structures were predicted using docking, energy funnels are shown in Figure 3 – figure supplement 1.

**Figure 4. Evolutionary couplings give accurate 3D structures of complexes**. EVcomplex predictions and comparison to crystal structure for (A) the methionine-importing transmembrane transporter heterocomplex MetNI from E. coli (PDB: 3tui) and (B) the gamma/epsilon subunit interaction of E. coli ATP synthase (PDB: 1fs0). Left panels: Complex contact map comparing predicted inter-ECs with EVcomplex score ≥ 0.8 (red dots, upper right quadrant) and intra-ECs (up to the last chosen inter-EC rank; green and blue dots, top left and lower right triangles) to close pairs in the complex crystal (dark/mid/light grey points for minimum atom distance cutoffs of 5/8/12 Å for inter-subunit contacts and dark/mid grey for 5/8 Å within the subunits). Inter-ECs with an EVcomplex score ≥ 0.8 are also displayed on the spatially separated subunits of the complex (red lines on green and blue cartoons, couplings closer than 8Å in solid red lines, dashed otherwise, lower left). Right panels: Superimposition of the top ranked model from 3D docking (green/blue cartoon, left) onto the complex crystal structure (grey cartoon), and close-up of the interface region with highly coupled residues (green/blue spheres).

**Figure 5. Evolutionary couplings in complexes of unknown 3D structure**. Inter-ECs for five de novo prediction candidates without E. coli or interaction homolog complex 3D structure (Subunits: blue/green cartoons; inter-ECs with EVcouplings score ≥ 0.8: red lines). For complex subunits which homomultimerize (light/dark green cartoon), inter-ECs are placed arbitrarily on either of the monomers to enable the identification of multiple interaction sites. Contact maps for all complexes with unsolved structures are provided in Figure 5 - figure supplement 1 and 2. Left to right: (1) the membrane subunit of methionine-importing transporter heterocomplex MetI (PDB: 3tui) together with its periplasmic binding protein MetQ (Swissmodel: P28635); (2) the large and small subunits of acetolactate synthase IlvB (Swissmodel: P08142) and IlvN (PDB: 2lvw); (3) panthotenate synthase PanC (PDB: 1iho) together with ketopantoate hydroxymethyltransferase PanB (PDB: 1m3v); (4) subunits a and b of ATP synthase (model for a subunit a predict with EVfold-membrane, PDB: 1b9u for b subunit), for detailed information see Figure 5; and (5) the in DNA repair and SOS mutagenisis involved complex UmuC (model created with EVfold) with one possible conformation of UmuD (PDB: 1i4v). For alternative UmuD conformation, see Figure 5 – figure supplement 3.

**Figure 6. Predicted interactions between the a-, b- and c- subunits of ATP synthase**. (A) The a- and b- subunits of E. coli ATP synthase are known to interact, but the monomer structure of subunits a and b and the structure of their interaction in the complex are unknown. (B) EVcomplex prediction

(right matrix) for ATP synthase subunit interactions compared to experimental evidence (left matrix), which is either strong (left, solid blue squares) or indicative (left, crosshatched squares). Interactions that have experimental evidence, but are not predicted at the 0.8 threshold are indicated as yellow dots. (C) Left panel: Residue detail of predicted residue-residue interactions (dotted lines) between subunit a and b (residue numbers at the boundaries of transmembrane helices in grey). Right panel: Proposed helix-helix interactions between ATP synthase subunits a (green), b (blue, homodimer), and the c ring (grey). The proposed structural arrangement is based on analysis of the full map of inter-subunit ECs with a EVcomplex score larger than 0.8 (Figure 6 - figure supplement 1).

Figure Supplements

Figure 1 – figure supplement 1: **Details of the EVcomplex Pipeline**

Figure 2 – figure supplement 1-2: **Distribution and accuracy of raw EC scores for all complexes in evaluation set**

Figure 2 – figure supplement 3-8: **Contact maps of all complexes with solved 3D structure with inter-ECs above EVcomplex score of 0.8.** Predicted coevolving residue pairs with an EVcomplex score ≥ 0.8 and all inter-ECs up to the rank of the last include inter-EC are visualized in complex contact maps (red dots: inter-ECs, green and blue dots: intra-ECs for monomer 1 and 2, respectively). Top left and bottom right quadrants: intra-ECs; top right and bottom left quadrants: inter-ECs. Inter- and intra-protein crystal structure contacts at minimum atom distance cutoffs of 5/8/12 Å are shown as dark/middle/light grey dots, respectively; missing data in the crystal structure as shaded blue rectangles.

Figure 3 – figure supplement 1: **Comparison of Interface RMSD to HADDOCK score.** The HADDOCK scores of docked models are plotted against their iRMSDs to the bound complex crystal. Grey data points correspond to models created without any ECs as unambiguous restraints whereas blue dots correspond to model created using all inter-couplings with EVcomplex score ≥ 0.8. HADDOCK score outliers with scores > 100 are not shown, and any model with an iRMSD > 35Å is displayed as iRMSD=35 Å for visualization purposes

Figure 5 – figure supplement 1-2: **Contact maps of all complexes without solved 3D structure with at least one inter-ECs above EVcomplex score of 0.8**. Inter-ECs are shown as red dots in the top right and bottom left quadrant while intra-ECs of the two monomers are shown in green and blue in the top left and bottom right quadrant, respectively.

Figure 5 – figure supplement 3: **Details of the predicted UmuCD interaction residues**

Figure 6 – figure supplement 1: **Contact map of predicted ECs in the ATPsynthase a and b subunits**. Inter-ECs are shown as red dots in the top right and bottom left quadrant while intra-ECs of the two monomers are shown in green and blue in the top left and bottom right quadrant, respectively.

Supplementary Files

Supplementary file 1: Benchmark dataset and results

Supplementary file 2: Unknowns dataset and results

Supplementary file 3: Docking results

Supplementary file 4: Predicted inter-ECs for complexes in de novo prediction dataset with EVcomplex score ≥ 0.8

Supplementary file 5: ATPsynthase predictions

Supplementary file 6: Comparison of ATP synthase EVcomplex predictions of a and b subunit with cross-linking studies

Supplementary file 7: PDB identifiers used for comparison of predicted evolutionary couplings to known 3D structures

Supplementary Data

Supplementary data 1: Concatenated alignments for complexes predicted in this work

Supplementary data 2: Genome distance distribution of concatenated sequences per alignment

Supplementary data 3: EVcomplex predictions for evaluation and *de novo* set

Supplementary data 4: Docking input files and top 10 predicted models for evaluation set

Supplementary data 5: ATP synthase predictions, ATP synthase subunit a model

Supplementary data 6: UmuC model

# Figure 1

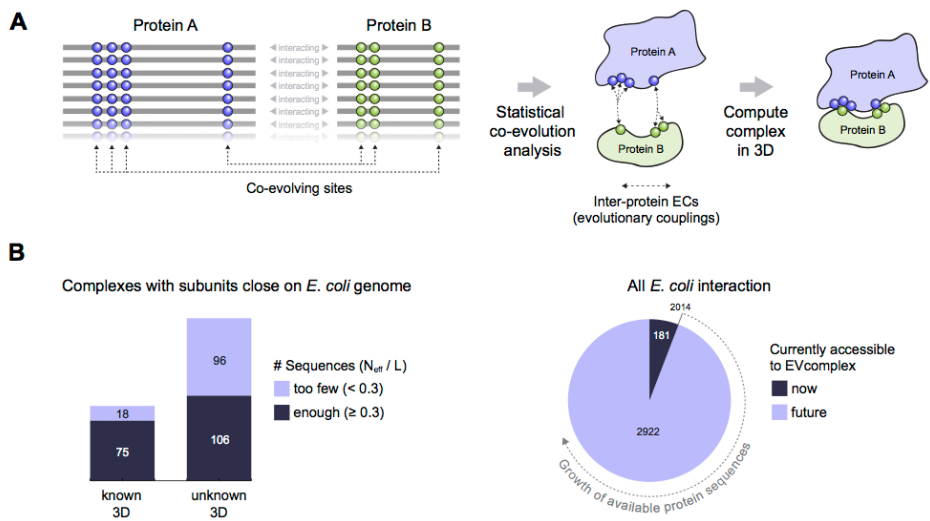

Figure 2

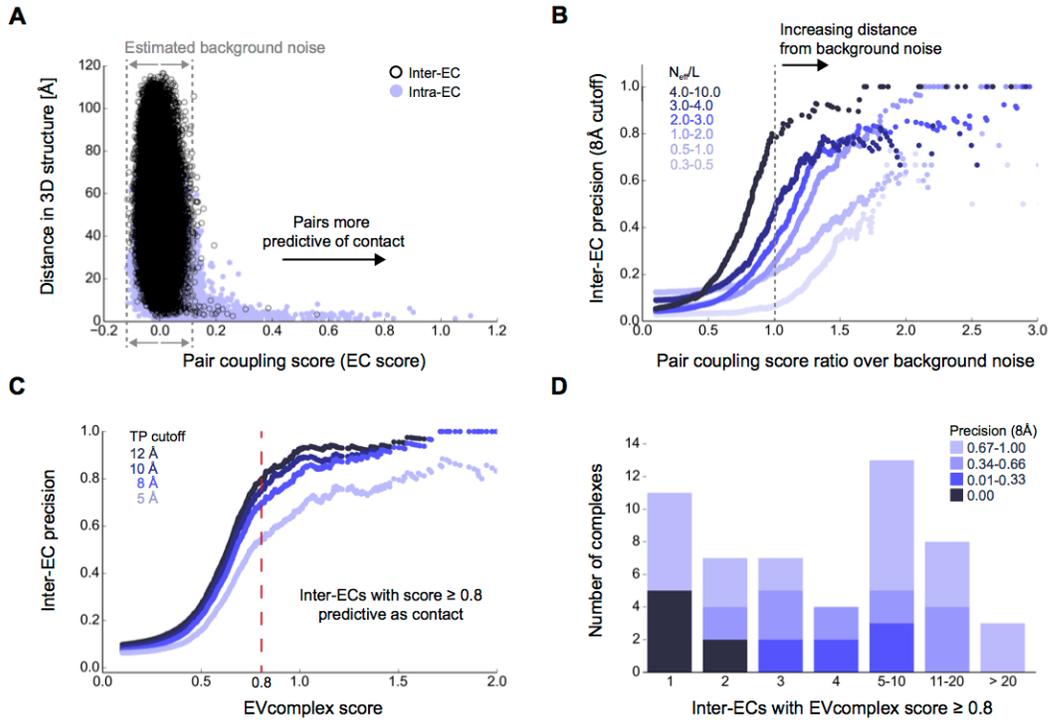

**Figure 3**

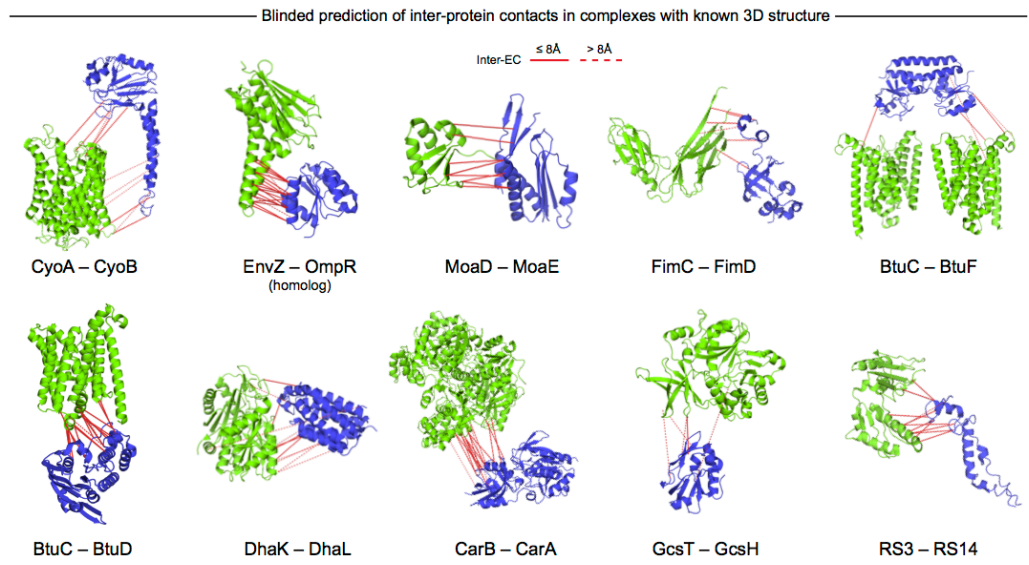

Blinded prediction of inter-protein contacts in complexes with known 3D structure

Inter-EC ≤ 8Å ——— > 8Å - - -

CyoA – CyoB

EnvZ – OmpR
(homolog)

MoaD – MoaE

FimC – FimD

BtuC – BtuF

BtuC – BtuD

DhaK – DhaL

CarB – CarA

GcsT – GcsH

RS3 – RS14

**Figure 4**

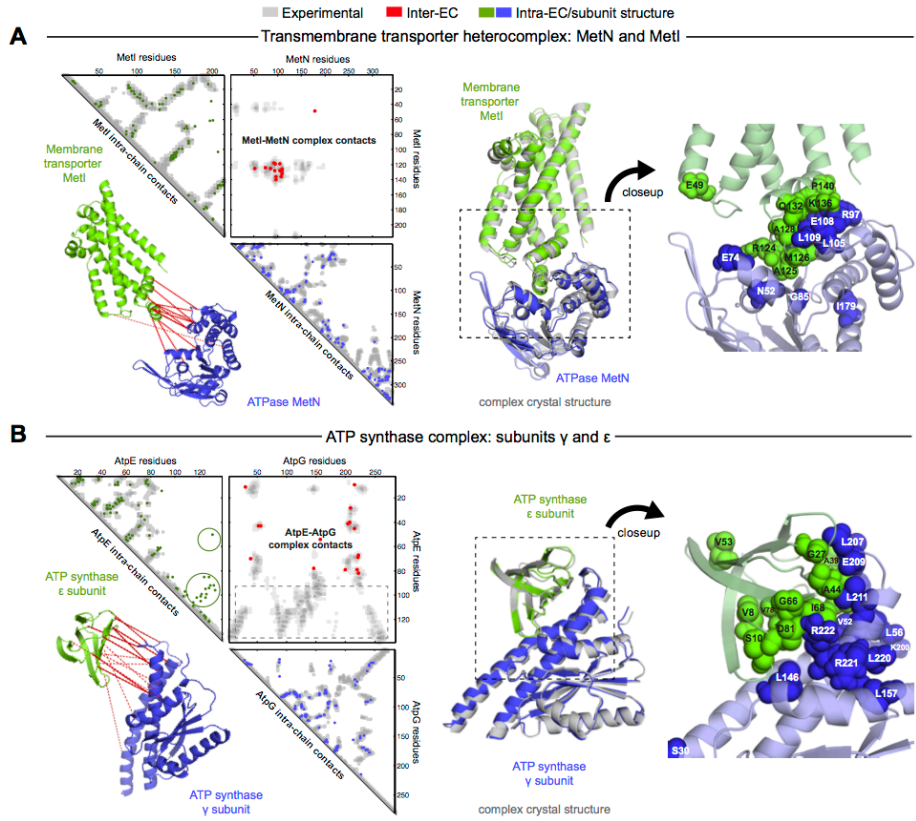

## Figure 5

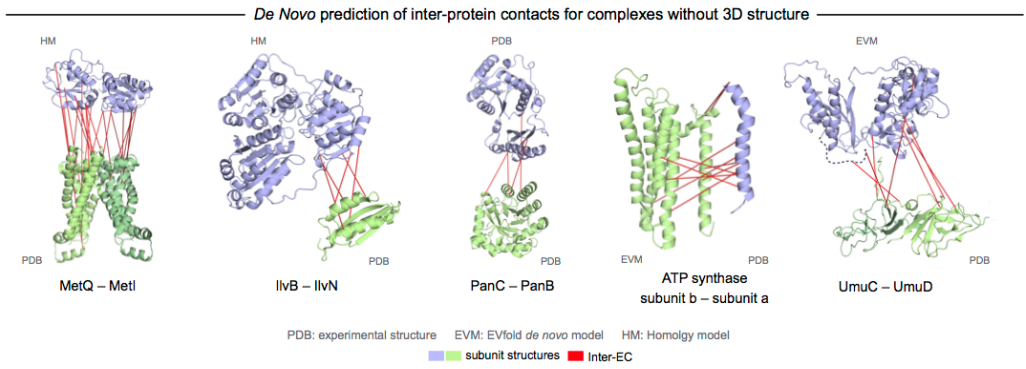

*De Novo* prediction of inter-protein contacts for complexes without 3D structure

MetQ – MetI   IlvB – IlvN   PanC – PanB   ATP synthase subunit b – subunit a   UmuC – UmuD

PDB: experimental structure   EVM: EVfold *de novo* model   HM: Homolgy model

subunit structures   Inter-EC

**Figure 6**

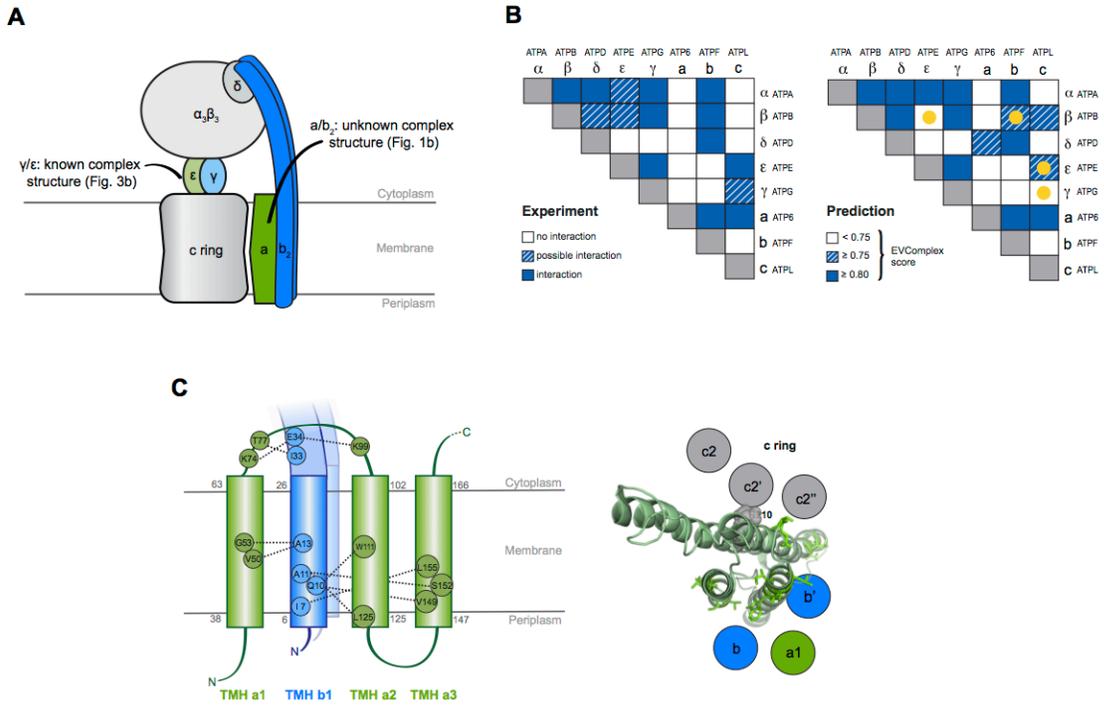

Prediction of the partially known complex of ATP synthase